\documentclass[intlimits,twoside,a4paper]{article}

\usepackage[cp1251]{inputenc}
\usepackage[eqsecnum]{cmpj3}
\articletype{A part of the Special collection to the 100th anniversary of the birth of Ihor Yukhnovskii}

\issue{2025}{28}{2}{23501}
\doinumber{10.5488/CMP.28.23501}

\title[A consistent description of the kinetic processes]%
      {A consistent description of the kinetic processes of electrolyte ion transport in a dynamic  porous medium}
\author[P. P. Kostrobij, B. M. Markovych, O. V. Viznovych, M. V. Tokarchuk]
       {P. P. Kostrobij\orcid{0000-0002-4428-1647}\refaddr{label1}, B. M. Markovych\orcid{0000-0002-8813-9108}\refaddr{label1}, O. V. Viznovych\orcid{0000-0002-5483-5113}\refaddr{label1}, M. V. Tokarchuk\orcid{0000-0002-9205-1790}\refaddr{label2,label1}}

\addresses{\addr{label1}Lviv Polytechnic National University, 12 S. Bandera Str., 79013 Lviv, Ukraine
           \addr{label2}Institute for Condensed Matter Physics of the National Academy of Sciences of Ukraine, 1 Svientsitskii Str., 79011 Lviv, Ukraine}
\date{Received April 08, 2025, in final form May 22, 2025}

\begin{document}
 \renewcommand{\vec}{\mathbf}
 \sloppy

\maketitle

\begin{abstract}
 The consistent description of kinetic and hydrodynamic processes is applied to the study of ion transport processes in the ionic solution--porous medium system.
 A system of equations is obtained for the nonequilibrium single--ion distribution function, the nonequilibrium average value of the energy density of the interaction of solution ions, and the nonequilibrium average value of the number density of particles in a porous medium.
 Using the fractional calculus technique, a generalized diffusion equation of the Cattaneo type in fractional derivatives is obtained to describe the processes of subdiffusion of particles in a porous medium.

\end{abstract}

 \section{Introduction}

 Studies of the structural and thermodynamic properties of charged particle, ionic, and ion-molecular systems, starting in 1954, were among the most important in I.~R.~Yukhnovskii's work based on the original method of collective variables~\cite{Yukhnovskii195967,Yukhnovskii1980,Yukhnovskii1988}.
 Such a statistical approach, which is still relevant~\cite{Yukhnovskii2022579}, made it possible to consistently describe the screening effects for ionic and ion-molecular systems and obtain analytical expressions for pair distribution functions in the approximations of the fourth group coefficients.
 Such pair distribution functions of ions, molecules (dipoles) were used for calculations in the appropriate approximations of diffusion coefficients and time correlation functions for ion-dipole systems~\cite{Tokarchuk19861123}, calculations of dynamic structural factors ``mass--mass'', ``mass--charge'', ``charge--charge'' for ionic melts NaCl within the framework of generalized hydrodynamics~\cite{Zubarev1987164}.
 Unary and pair distribution functions of ions and molecules in spatially inhomogeneous systems obtained in the collective variable method~\cite{Yukhnovskii1988} were applied in the studies of transfer (diffusion) coefficients in spatially inhomogeneous systems:
 ``electrolyte -- membrane -- filtrate'' (modelling of reverse osmotic processes of purification of aqueous solutions)~\cite{Yukhnovskii1996819},
 ``aqueous solutions of radionuclides -- soil'' (modelling of radionuclide migration processes in groundwater in the Chernobyl zone)~\cite{Yukhnovskii2000361}, ``aqueous solutions of radionuclides -- fuel-containing masses'' (modelling of the interaction processes of aqueous solutions of radionuclides with fuel-containing masses in the New Safe Confinement of the ``Shelter'' object in the Chernobyl zone)~\cite{Yukhnovskii1999351,Tokarchuk20251}.
 It is important to note that reverse osmosis membranes, soil, and fuel-containing masses are porous systems that interact with aqueous solutions containing ions of different types with different properties.

 Mathematical modelling of ion transport in porous systems from the point of view of modern technologies for creating new current sources, water purification devices, biological membranes, channels, etc.~\cite{Vasylenko2024,Kondori2023,Yu2024,Ding2024,Biesheuvel2019,Yang2019,Heiranian2023,Fan2023,Miao2024,Ye2024,Stabilini2021,Su2024144} remains relevant. This direction of research allows for a deeper understanding of these processes and their control.

 It is important to use the statistical theory of diffusion processes in the ``electrolyte -- porous medium'' system with equal consideration of both the electrolyte and the porous medium.
 In the vast majority of studies, non-equilibrium thermodynamic equations~\cite{Ferguson20121967} with constant diffusion coefficients are used to describe electrodiffusion processes of ions, in particular in systems ``electrolyte--electrode''.
 At the same time, an important feature of these systems is their significant spatial heterogeneity, when the diffusion coefficients are functions of spatial coordinates and time, i.e., the time correlation functions ``flow--flow'' $\langle\hat{\vec{j}}(\vec{r}_{l};t)\hat{\vec{j}}(\vec{r}_{l'};t')\rangle$ in each of the phases and between the phases.
 In modern conditions of using current batteries, one of the important processes is a rapid charging of the battery electrodes with the corresponding charge carriers.
 In this case, we have a situation where, as the corresponding charge carriers accumulate, the structure and properties of the storage material (electrode) will change accordingly, in particular the porosity of the material.
 A change in the structure and porosity leads to a change in the resistance of the storage material.
 It increases, and further accumulation by ions of the material becomes more problematic due to polarization, diffusion and structural processes in the storage material itself.
 Obviously, there must be a certain balance between the possibility of accumulation of charge carriers and the applied charging voltage.
 In this regard, it is necessary to equally take into account non-equilibrium processes for particles for the electrolyte and the porous medium.

 A statistical theory for describing electrodiffusion processes of ion transport in the ``electrolyte -- electrode'' system, taking into account spatial heterogeneity and memory effects, was proposed in the works~\cite{Kostrobij2014177,Grygorchak2015e,Kostrobij2015154,Grygorchak2017185501,Kostrobij2017163,Kostrobij20184099}, using the method of non-equilibrium statistical operator (NSO).
 In particular, experimental and theoretical~\cite{Grygorchak2015e,Kostrobij2015154,Grygorchak2017185501,Kostrobij20184099} studies of subdiffusion impedance for a multilayer GaSe system with encapsulated $\beta$-cyclodextrin, which has a fractal porous structure, were carried out.
 From the point of view of theoretical studies, generalized electrodiffusion equations of the Cattaneo-type in fractional derivatives~\cite{Kostrobij20184099} were applied.

 In a recent work~\cite{Tokarchuk20241013}, a kinetic approach was applied to describe ion transport processes in the ionic solution-porous medium system based on the Enskog--Landau kinetic equations~\cite{Kobryn1996189} for ions, which are obtained from a chain of BBGKI equations with modified boundary conditions.
 By solving the Enskog--Landau kinetic equations for charged solid spheres and constructing hydrodynamic equations, analytical expressions for the coefficients of mutual diffusion, thermal diffusion, viscosity, and thermal conductivity were obtained for stationary processes, taking into account the porosity of the medium.

 In this work, the kinetic approach is applied to the description of ion transport processes in the system ``ionic solution -- porous medium''.
 In the second section, the non-equilibrium particle distribution function for the system ``ionic solution -- porous medium'' is obtained using the Zubarev's method of the non-equilibrium statistical operator~\cite{Zubarev19811509,Zubarev20021,Zubarev20022}.
 The non-equilibrium properties of the ionic solution are described within the framework of the agreement of kinetics and hydrodynamics, and the particles of the porous medium within the framework of the diffusion (subdiffusion) description.
 A system of equations is obtained for the non-equilibrium single-ion distribution function, the non-equilibrium average value of the energy density of the interaction of solution ions (taking into account the interaction with the particles of the porous medium) and for the non-equilibrium average value of the number density of the particles of the porous medium.
 In the third section, using the fractional calculus technique~\cite{Samko1993,Oldham2006}, a generalized diffusion equation of the Cattaneo-type in fractional derivatives is obtained to describe the processes of subdiffusion of particles in a porous medium.

 \section{Non-equilibrium statistical operator for the ionic solution-porous medium system}

 We consider a system of an ionic solution that interacts with a porous medium, diffusing in it.
 Positively and negatively charged ions of the solution can penetrate the structure of the porous medium and move in it, interacting with particles of the porous medium (matrix).
 We take the entire volume of the system to be equal to $V=V_{l}+V_{s}$, where $V_{l}$ is the volume occupied by the ionic solution, and $V_{s}$ is the true volume of the porous matrix.
 By introducing the volume $V_{\rm por}$ of the porous space of the matrix, we can determine its porosity: $\phi={V_{\rm por}}/{V}=1-{V_{s}}/{V}$.
 We consider the ionic solution with certain dielectric properties without explicit consideration of the molecular subsystem, and the porous matrix as formed by structureless moving particles of the $\xi$ species.
 Examples of a porous medium can be biological systems (macromolecular structures), polymeric, composite materials, electrode materials, which in the processes of their interaction with ionic solutions can change structurally, which is extremely important to take into account.
 Unlike an ionic solution, a porous medium as a liquid can be considered as an amorphous or partially amorphous structural system, the particles of which can diffuse due to structural displacements, changing the structure of the material over a long time.
 In reality, electrode, membrane structures in the processes, in particular, ion transport in them, undergo significant changes due to temperature, polarization and other factors.
 Such changes can occur very slowly by subdiffusion of particles of the porous medium due to interaction with, in particular, in our case, an ionic solution during a long relaxation time.
 Therefore, we assume that the particles of the porous matrix can perform very slow diffusional motion for a long time of the subdiffusion type~\cite{Sandev201810} $\langle (\vec{R}(t)-\vec{R}(0))^{2}\rangle=2D_{\nu}t^{\nu}[{1}/{\Gamma(1+\nu)}]$, where $D_{\nu}$ is the coefficient of ``anomalous'' diffusion of particles of the porous medium, $\vec{R}(t)$ are the coordinates of the particles, $t$ is time, $\Gamma(1+\nu)$ is the gamma function, the parameter $\nu$ varies in the interval $0<\nu\leqslant 1$.
 The reason for such subdiffusion processes can be the interaction of particles of the porous medium with solution ions, the change in the pore space when they are filled with an ionic solution, and other factors.
 Considering the system in this way, we have $f_{\xi}=1-\phi$ --- the volume fraction of the porous matrix, $f_{l}=s_{l}\phi$ --- the volume fraction of the pores filled with an ionic solution, where $s_{l}$ is the degree of filling the pores with an ionic solution (at $s_{l}=1$ is complete saturation of the porous medium with ionic solutions) and $f_{n}=(1-s_{l})\phi$ is the volume fraction of incomplete filling of the pores with an ionic solution.
 We assume that the kinetic energy of the solution ions is much greater than the kinetic energy of the particles of the porous medium.

 The Hamiltonian of such a system, an ionic solution -- a porous matrix, can be represented as:
 \begin{align}\label{e.01}
  H&=\sum_{\alpha,j=1}^{N_{\alpha}}\frac{\left[\vec{p}_{j}-\frac{Z_{\alpha}e}{c}\vec{A}(\vec{r}_{j};t)\right]^{2}}{2m_{\alpha}} +\sum_{\alpha,\gamma}\sum_{j,l}^{N_{\alpha},N_{\gamma}}\Phi_{\alpha\gamma}(\vec{r}_{j},\vec{r}_{l}) \nonumber\\
  &+\sum_{\alpha,\xi}\sum_{j,l}^{N_{\alpha},N_{\xi}}\Phi_{\alpha \xi}(\vec{r}_{j},\vec{R}_{l})
        +\sum_{\alpha}\sum_{j}^{N_{\alpha}}Z_{\alpha}e\varphi(\vec{r}_{j};t)+H_{\xi},
 \end{align}
 where the indices $j$, $l$ number the solution ions of the type $\alpha$, $\gamma$,
 with masses $m_{\alpha}$, $m_{\gamma}$ and $\vec{p}_{j}$, $\vec{r}_{j}$ are the momentum vector and the coordinate vector of the $j$-th ion of the solution, $\vec{P}_{l}$, $\vec{R}_{l}$ are the momentum vector and the coordinate vector of the $l$-th particle of the porous matrix, $N_{\alpha}$ is the total number of ions of the type $\alpha$;
 $\Phi_{\alpha\gamma}(\vec{r}_{j},\vec{r}_{l})=\Phi^{\rm sh}_{\alpha\gamma}(\vec{r}_{j},\vec{r}_{l})+\Phi^{\rm long}_{\alpha\gamma}(\vec{r}_{j},\vec{r}_{l})$ is the pair potential of interaction between ions of the type $\alpha,\gamma$;
 $\Phi_{\alpha \xi}(\vec{r}_{j},\vec{R}_{l})=\Phi^{\rm sh}_{\alpha \xi}(\vec{r}_{j},\vec{R}_{l})+\Phi^{\rm long}_{\alpha \xi}(\vec{r}_{j},\vec{R}_{l})$ is the pair potential of interaction of ions with particles of the porous matrix, which have short-range and long-range contributions.
 $\vec{A}(\vec{r}_{j};t)$, $\varphi(\vec{r}_{j};t)$ are the total vector and scalar potentials of the electromagnetic field created by ions of valence $Z_{\alpha}$ and the external field, $e$ is the electron charge and $c$ is the speed of light.
 In what follows, we do not consider vortex electromagnetic processes of order $\frac{e}{c}$, but only potential contributions from $\varphi(\vec{r}_{j};t)$.
 $H_{\xi}$ is the Hamiltonian of the particles of the porous matrix, which includes the kinetic and potential parts.
 In addition, at this stage of the research we do not take into account the possible influence of the electromagnetic field on the particles of the porous matrix, which may be associated with the processes of polarization and changes in its dielectric properties.
 Although in reality, in many cases, in particular for electrodes (for instance, dipole superparamagnets of the type $\langle\beta$-CD$\langle$FeSO$_4\rangle\rangle$ in semiconductor layered structures, such as GaSe, InSe \cite{Grygorchak2017185501,Kostrobij20184099,Klapchuk2020322,Kostrobij202189}), biological membranes, and macromolecular structures, such consideration is important.

 The non-equilibrium state of an ionic solution, when interacting with a porous matrix, is completely described by the Liouville equation for the non-equilibrium distribution function of all particles $\rho(x_{1},\ldots,x_{N_{\alpha}}|X_{1},\ldots,X_{N_{\xi}};t)=\rho(x^{N_{\alpha}},X^{N_{\xi}};t)$ (where the notations are introduced: $x^{N_{\alpha}}=x_{1},\ldots,x_{N_{\alpha}}$, $x_{j}=\mathbf p_j,\,\mathbf r_j$, $X^{N_{\xi}}=X_{1},\ldots,X_{N_{\xi}}$, $X_{1}=\vec{P}_{l},\,\vec{R}_{l}$)
 \begin{equation}\label{e.001}
  \frac{\partial }{\partial t}\rho\left(x^{N_{\alpha}},X^{N_{\xi}};t\right)+\ri L_{N}\rho\left(x^{N_{\alpha}},X^{N_{\xi}};t\right)=0,
 \end{equation}
 with Liouville operator
 \begin{align}\label{e.02}
  \ri L_{N}(t) &=\sum_{\alpha,j=1}^{N_{\alpha}}\frac{\vec{p}_{j}}{m_{\alpha}}\cdot \frac{\partial }{\partial \vec{r}_{j}}
  -\sum_{\alpha,\gamma}\sum_{j,l}^{N_{\alpha}N_{\gamma}}\frac{\partial }{\partial \vec{r}_{j}}\Phi_{\alpha\gamma}(\vec{r}_{j},\vec{r}_{l})
  \left(\frac{\partial }{\partial \vec{p}_{j}}- \frac{\partial }{\partial \vec{p}_{l}}\right) \nonumber\\
  &-\sum_{\alpha,\xi}\sum_{j,l}^{N_{\alpha},N_{\xi}}\frac{\partial}{\partial \vec{r}_{j}}
   \big[\Phi_{\alpha \xi}(\vec{r}_{j},\vec{R}_{l})+Z_{\alpha}e\varphi(\vec{r}_{j};t)\big]\cdot \frac{\partial }{\partial \vec{p}_{j}} \nonumber\\
  &+ \sum_{\xi,l=1}^{N_{\xi}}\frac{\vec{P}_{l}}{m_{\xi}}\cdot \frac{\partial }{\partial \vec{R}_{l}}
   -\sum_{\xi,\xi'}\sum_{l,l'}^{N_{\xi}N_{\xi'}}\frac{\partial }{\partial \vec{R}_{l}}\Phi_{\xi\xi'}(\vec{R}_{l},\vec{R}_{l'})
   \left(\frac{\partial }{\partial \vec{P}_{l}}-\frac{\partial }{\partial \vec{P}_{l'}}\right) \nonumber\\
  &-\sum_{\xi,\alpha}\sum_{l,j}^{N_{\alpha},N_{\xi}}\frac{\partial }{\partial \vec{R}_{l}}
  \Phi_{\xi \alpha}(\vec{R}_{l},\vec{r}_{j})\cdot \frac{\partial }{\partial \vec{P}_{l}}.
 \end{align}
 To find the non-equilibrium distribution function $\rho(x^{N_{\alpha}},X^{N_{\xi}};t)$ we use the approach proposed in the works~\cite{Zubarev1991412,Zubarev1993997,Tokarchuk1998687,Tokarchuk2022440} of a consistent description of the kinetics and hydrodynamics of non-equilibrium processes of a system of interacting particles in  the Zubarev's method of the non-equilibrium statistical operator, based on the Liouville equation with the source:
 \begin{equation}\label{e.0010}
  \frac{\partial}{\partial t}\rho\left(x^{N_{\alpha}},X^{N_{\xi}};t\right)
  +\ri L_{N}\rho\left(x^{N_{\alpha}},X^{N_{\xi}};t\right)
  =
  -\varepsilon\left(\rho\left(x^{N_{\alpha}},X^{N_{\xi}};t\right)-\varrho_{\rm rel}\left(x^{N_{\alpha}},X^{N_{\xi}};t\right)\right),
 \end{equation}
 which selects the retarded ($\varepsilon\rightarrow +0$, after the thermodynamic transition) solutions of the Liouville equation under given initial conditions.
 As such an initial condition for the solution of the Liouville equation (Cauchy problem) we choose:
 \[
  \rho\left(x^{N_{\alpha}},X^{N_{\xi}};t\right)_{t=t_{0}}
  =
  \varrho_{\rm rel}\left(x^{N_{\alpha}},X^{N_{\xi}};t_{0}\right),
 \]
 when the ionic solution and the porous matrix are considered as interacting subsystems at the initial time.
 In this case, $\rho_{\rm rel}\left(x^{N_{\alpha}},X^{N_{\xi}};t\right)$ is the relevant distribution function of ions and particles of the porous matrix, which is obtained according to~\cite{Zubarev1991412,Zubarev1993997,Tokarchuk1998687,Tokarchuk2022440} from the condition of the maximum of the Gibbs entropy functional while maintaining the normalization conditions for the distribution and the given parameters of the abbreviated description of the non-equilibrium state of the ionic solution; $\langle\hat{n}_{\alpha}(x)\rangle^{t}=f_{\alpha}(x;t)$ is the non-equilibrium single-particle distribution function of ion species $\alpha$, $\langle \hat{\varepsilon}_{\rm int}(\vec{r})\rangle^{t}$ is the non-equilibrium average energy of interaction of solution ions and particles of the porous matrix, $\langle\hat{n}_{\xi}(\vec{R})\rangle^{t}$ is the non-equilibrium average value of the density of particles  species $\xi$ of the porous matrix  and has the following structure:
 \begin{align}\label{e.0021}
  \rho_{\rm rel}\left(x^{N_{\alpha}},X^{N_{\xi}};t\right)
  &=
  \exp\bigg[-\Phi(t)-\int \rd\vec{r}\,\beta(\vec{r},t)\,\hat{\varepsilon}_{\rm int}(\vec{r})
  -\sum_{\alpha}\int \rd x \,a_{\alpha}(x,t)\,\hat{n}_{\alpha}(x) \nonumber\\
  &-\beta\bigg(H_{\xi}-\sum_{\xi}\int \rd\vec{R}\,\mu_{\xi}(\vec{R};t)\,\hat{n}_{\xi}(\vec{R})\bigg)\bigg],
 \end{align}
 where $\Phi(t)$ is the Massier--Planck functional:
 \begin{align}\label{e.0031}
  \Phi(t)=\int \rd\Gamma(x,X)
  \exp\bigg[&-\int \rd\vec{r}\,\beta(\vec{r},t)\,\hat{\varepsilon}_{\rm int}(\vec{r})
  -\sum_{\alpha}\int \rd x \, a_{\alpha}(x,t)\,\hat{n}_{\alpha}(x) \nonumber\\
  &-\beta\bigg(H_{\xi}-\sum_{\xi}\int \rd\vec{R}\,\mu_{\xi}(\vec{R};t)\,\hat{n}_{\xi}(\vec{R})\bigg)\bigg].
 \end{align}
 Here, $\hat{n}_{\alpha}(x)$ is the microscopic phase number density of ion  species  $\alpha$:
 \begin{equation}\label{e.004aa}
 \hat{n}_{\alpha}(x)=\sum_{j=1}^{N_{\alpha}}\delta(\vec{p}-\vec{p}_{j})\delta(\vec{r}-\vec{r}_{j}),
 \end{equation}
 \begin{equation}\label{e.004a}
  \hat{n}_{\xi}(\vec{R})=\sum_{l=1}^{N_{\xi}}\delta (\vec{R}-\vec{R}_{l})
 \end{equation}
is the microscopic density of particles of the porous medium,  and
 \begin{equation}       \label{e.0051}
  \hat{\varepsilon}_{\rm int}(\vec{r})=\frac{1}{2}\sum_{\alpha,\gamma}\sum_{j,l}^{N_{\alpha},N_{\gamma}}
  \Phi_{\alpha\gamma}(\vec{r}_{j},\vec{r}_{l}) \,\delta (\vec{r}-\vec{r}_{j})
  +\sum_{\alpha}\sum_{j,l}^{N_{\alpha},N_{\xi}}
  \Phi_{\alpha \xi}(\vec{r}_{j},\vec{R}_{l}) \,\delta(\vec{r}-\vec{r}_{j})
 \end{equation}
 is the microscopic energy density of the interaction between solution ions and porous matrix particles.
 The Lagrangian parameters $\beta(\vec{r},t)$ (the inverse of the non-equilibrium temperature of the system ``ionic solution -- porous matrix''), $a_{\alpha}(x,t)$ and $\mu_{\xi}(\vec{R},t)$ are determined from the self-consistency conditions:
 \begin{gather}\label{e.0014}
  \langle \hat{\varepsilon}_{\rm int}(\vec{r})\rangle^{t}=\langle \hat{\varepsilon}_{\rm int}(\vec{r})\rangle^{t}_{\rm rel}, \quad
  \langle\hat{n}_{\alpha}(x)\rangle^{t}= \langle\hat{n}_{\alpha}(x)\rangle^{t}_{\rm rel}, \\
  \langle\hat{n}_{\xi}(\vec{R})\rangle^{t}= \langle\hat{n}_{\xi}(\vec{R})\rangle^{t}_{\rm rel}. \nonumber
 \end{gather}
 In this case, the ionic solution has a temperature of $\beta^{-1}(\vec{r},t)$, and $\beta^{-1}$ is the temperature of the porous matrix.
 $\langle\ldots\rangle^{t}_{\rm rel}=\int \rd\Gamma_{N} \ldots \varrho_{\rm rel}\big(x^{N_{\alpha}},X^{N_{\xi}};t\big)$,
 $\langle\ldots\rangle^{t}=\int \rd\Gamma_{N} \ldots \varrho\big(x^{N_{\alpha}},X^{N_{\xi}};t\big)$, $\rd\Gamma_{N}=\frac{(\rd x_{1}\ldots{\rd}x_{N_{\alpha}})}{N_{\alpha}!}\frac{(\rd X_{1}\ldots \rd X_{N_{\xi}})}{N_{\xi}!}$,
 $\rd x=\rd{\vec r}\,\rd{\vec p}$,
 $\rd X=\rd{\vec R}\,\rd{\vec P}$.

 Next, we use the general solution of the Liouville equation taking into account the design procedure in the NSO method, which can be presented in the form:
 \begin{align}\label{eq2.6}
  \varrho\left(x^{N_{\alpha}},X^{N_{\xi}};t\right)
 &=
  \varrho_{\rm rel}\left(x^{N_{\alpha}},X^{N_{\xi}};t\right) \nonumber\\
  &-
  \int_{-\infty}^{t} \rd t' \, \re^{\epsilon(t'-t)}\,T_{\rm rel}(t,t')\,
  \big[1-P_{\rm rel}(t')\big]\,\ri L_{N}\varrho_{\rm rel}\left(x^{N_{\alpha}},X^{N_{\xi}};t'\right),
 \end{align}
 where $\epsilon\rightarrow +0$ after the limiting thermodynamic transition, which selects the retarded solutions of the Liouville equation with the operator $\ri L_{N}$.
 $T_{\rm rel}(t,t')=\exp_{+}\big\{-\int_{t'}^{t}\rd t'[1-P_{\rm rel}(t')]\ri L_{N}\big\}$ is a generalized time-dependent evolution operator taking into account the Kawasaki--Gunton projection $P_{\rm rel}(t)$.
 The structure of $P_{\rm rel}(t)$ depends on the relevant distribution function $\varrho_{\rm rel}\big(x^{N_{\alpha}},X^{N_{\xi}};t\big)$.
 Having revealed the action of the operators $(1-P_{\rm rel}(t))$ and $\ri L_{N}$ on the relevant distribution function $\varrho_{\rm rel}\big(x^{N_{\alpha}},X^{N_{\xi}};t'\big)$,
 for $\varrho\big(x^{N_{\alpha}},X^{N_{\xi}};t\big)$ we obtain
 \begin{align}    \label{eq22.88}
  \varrho(x^{N_{\alpha}},X^{N_{\xi}};t)&=\varrho_{\rm rel}(x^{N_{\alpha}},X^{N_{\xi}};t) \nonumber\\ 
  &+\bigg\{\sum_{\gamma}\int \rd x'\int_{-\infty}^{t}
  \re^{\epsilon(t'-t)}T_{\rm rel}(t,t')\,[1-P(t')]\,\ri L_{N}\hat{n}_{\gamma}(x')a_{\gamma}(x',t') \nonumber\\
  &+\int \rd\vec{r}'\int_{-\infty}^{t}
  \re^{\epsilon(t'-t)}T_{\rm rel}(t,t')\,[1-P(t')]\,\ri L_{N}\hat{\varepsilon}_{\rm int}(\vec{r}')\beta (\vec{r}',t') \nonumber\\
  &+\int \rd\vec{R}'\int_{-\infty}^{t}
  \re^{\epsilon(t'-t)}T_{\rm rel}(t,t')\,[1-P(t')]\,\ri L_{N}\hat{n}_{\xi}(\vec{R}')\beta \mu_{\xi}(\vec{R}',t')\bigg\}\nonumber\\
  &\times \varrho_{\rm rel}\big(x^{N_{\alpha}},X^{N_{\xi}};t'\big)\,\rd t',
 \end{align}
 where $P(t)$ is the generalized Mori operator, which is related to the Kawasaki--Gunton operator:
 \[
  P_{\rm rel}(t)A\varrho_{\rm rel}(t)=\varrho_{\rm rel}(t)P(t)A
 \]
 and has the following structure:
 \begin{align*}
  P(t)A &=  \langle A \rangle_{\rm rel}^{t}
  +\sum_{\alpha}\int \rd\vec{r}\,\frac{\delta\langle A\rangle_{\rm rel}^{t}}{\delta\langle\hat{n}_{\alpha}(x)\rangle^{t}}
  \big[\hat{n}_{\alpha}(x)-\langle\hat{n}_{\alpha}(x)\rangle^{t}\big] 
  +\int \rd\vec{r}\,\frac{\delta \langle A\rangle_{\rm rel}^{t}}{\delta\langle\hat{\varepsilon}_{\rm int}(\vec{r})\rangle^{t}}
  \big[\hat{\varepsilon}_{\rm int}(\vec{r})-\langle \hat{\varepsilon}_{\rm int}(\vec{r})\rangle^{t}\big] \\
  &+\int \rd\vec{R}\,\frac{\delta \langle A \rangle_{\rm rel}^{t}}{\delta\langle\hat{n}_{\xi}(\vec{R})\rangle^{t}}
  \big[\hat{n}_{\xi}(\vec{R})-\langle \hat{n}_{\xi}(\vec{R})\rangle^{t}\big].
 \end{align*}
 The non-equilibrium distribution function~(\ref{eq22.88}) contains the generalized fluxes $[1-P(t')]\ri L_{N}\hat{n}_{\gamma}(x')$,
 $[1-P(t')]\ri L_{N}\hat{\varepsilon}_{\rm int}(\vec{r}')$,
 $[1-P(t')]\ri L_{N}\hat{n}_{\xi}(\vec{R}')$,
 as well as the Lagrangian parameters $a_{\gamma}(x',t')$, $\beta (\vec{r}',t')$, $\beta \mu_{\xi}(\vec{R}',t')$
 [determined from the self-consistency conditions~(\ref{e.0014})],
 which describe the corresponding non-Markovian transport processes in the system.
 Using the non-equilibrium distribution function for the parameters of the abbreviated description of non-equilibrium processes
 $\langle\hat{n}_{\alpha}(x)\rangle^{t}$,
 $\langle\hat{\varepsilon}_{\rm int}(\vec{r})\rangle^{t}$,
 $\delta\langle\hat{n}_{\xi}(\vec{R})\rangle^{t}$,
 using the following relation:
 \begin{align*}
  \frac{\partial }{\partial t}\langle \hat{n}_{\alpha}(x) \rangle^{t}&=\langle \dot{\hat{n}}_{\alpha}(x) \rangle^{t}_{\rm rel}+\langle (1-P(t)) \dot{\hat{n}}_{\alpha}(x) \rangle^{t}, \\
  \frac{\partial }{\partial t}\langle \hat{\varepsilon}_{\rm int}(\vec{r}) \rangle^{t}&=\langle \dot{\hat{\varepsilon}}_{\rm int}(\vec{r}) \rangle^{t}_{\rm rel}+
  \langle (1-P(t)) \dot{\hat{\varepsilon}}_{\rm int}(\vec{r}) \rangle^{t}, \\
  \frac{\partial }{\partial t}\langle \hat{n}_{\xi}(\vec{R}) \rangle^{t}&=\langle \dot{\hat{n}}_{\xi}(\vec{R}) \rangle^{t}_{\rm rel}+\langle (1-P(t))\dot{\hat{n}}_{\xi}(\vec{R}) \rangle^{t},
 \end{align*}
 we obtain a system of equations for ion transport in the system ``ionic solution -- porous medium'':
 \begin{align}\label{eq22.9}
  \frac{\partial}{\partial t}\big\langle\hat{n}_{\alpha}(x)\big\rangle^{t}
  &=\langle\dot{\hat{n}}_{\alpha}(x) \rangle^{t}_{\rm rel}
  +\sum_{\gamma}\int \rd x'\int_{-\infty}^{t}\re^{\varepsilon(t'-t)}\varphi_{nn}^{\alpha\gamma}(x,x';t,t')\,a_{\gamma}(x',t')\,\rd t' \nonumber\\
  &+\int \rd\vec{r}'\int_{-\infty}^{t}\re^{\varepsilon (t'-t)}\varphi_{n\varepsilon}^{\alpha}(x,\vec{r}';t,t')\,\beta (\vec{r}',t')\,\rd t' \nonumber\\
  &+\int \rd\vec{R}'\int_{-\infty}^{t}\re^{\varepsilon (t'-t)}\varphi_{nn}^{\alpha\xi}(x,\vec{R}';t,t')\,\beta \mu_{\xi}(\vec{R}',t')\,\rd t' ,
 \end{align}
 \begin{align}    \label{eq22.9a}
  \frac{\partial }{\partial t}\langle \hat{\varepsilon}_{\rm int}(\vec{r}) \rangle^{t}&=\langle \dot{\hat{\varepsilon}}_{\rm int}(\vec{r}) \rangle^{t}_{\rm rel}
  +\sum_{\gamma}\int \rd x'\int_{-\infty}^{t}\re^{\varepsilon (t'-t)}\varphi_{\varepsilon n}^{\gamma}(\vec{r},x';t,t')\,a_{\gamma}(x',t')\,\rd t' \nonumber\\
  &+\int \rd\vec{r}'\int_{-\infty}^{t}\re^{\varepsilon (t'-t)}\varphi_{\varepsilon\varepsilon}(\vec{r},\vec{r}';t,t')\,\beta(\vec{r}',t')\,\rd t' \nonumber\\
  &+\int \rd\vec{R}'\int_{-\infty}^{t}\re^{\varepsilon (t'-t)}\varphi_{\varepsilon n}^{\xi}(\vec{r},\vec{R}';t,t')\,\beta\mu_{\xi}(\vec{R}',t')\,\rd t',
 \end{align}
 \begin{align}\label{eq22.9b}
  \frac{\partial }{\partial t}\langle \hat{n}_{\xi}(\vec{R}) \rangle^{t}&=
  \sum_{\gamma}\int \rd x'\int_{-\infty}^{t}\re^{\varepsilon (t'-t)}\varphi_{nn}^{\xi\gamma}(\vec{R},x';t,t')\,a_{\gamma}(x',t')\,\rd t' \nonumber\\
  &+\int \rd\vec{r}'\int_{-\infty}^{t}\re^{\varepsilon (t'-t)}\varphi_{n\varepsilon}^{\xi}(\vec{R},\vec{r}';t,t')\,\beta(\vec{r}',t')\,\rd t' \nonumber\\
  &+\int \rd\vec{R}'\int_{-\infty}^{t}\re^{\varepsilon (t'-t)}\varphi_{nn}^{\xi\xi}(\vec{R},\vec{R}';t,t')\,\beta\mu_{\xi}(\vec{R}',t')\,\rd t',
 \end{align}
 where
 \begin{equation}\label{eq22.10}
  \dot{\hat{n}}_{\alpha}(x)
  =
  \ri L_{N}\hat{n}_{\alpha}(x)
  =
  -\frac{1}{m_{\alpha}}\frac{\partial }{\partial \vec{r}}\cdot \hat{\vec{j}}_{\alpha}(\vec{p},\vec{r})
  -\frac{\partial}{\partial \vec{p}}\cdot
  \bigg[\sum_{\gamma}\vec{F}_{\alpha\gamma}(\vec{p},\vec{r})+ \vec{F}_{\alpha\xi}(\vec{p},\vec{r})+Z_{\alpha}e\vec{E}(\vec{p},\vec{r};t)\bigg],
 \end{equation}
 \[
  \hat{\vec{j}}_{\alpha}(\vec{p},\vec{r})=\sum_{j=1}^{N_{\alpha}}\vec{p}_{j}\delta(\vec{p}-\vec{p}_{j})\delta(\vec{r}-\vec{r}_{j})
 \]
 is the microscopic momentum density of ion species $\alpha$ in the space of momentum and coordinates,
 and $\int \rd\vec{p}\,\hat{\vec{j}}_{\alpha}(\vec{p},\vec{r})=\hat{\vec{p}}_{\alpha}(\vec{r})$ is the microscopic momentum density of ion species $\alpha$.
 \begin{align*}
  \vec{F}_{\alpha\gamma}(\vec{p},\vec{r})&=\sum_{j,l}^{N_{\alpha}N_{\gamma}}\frac{\partial }{\partial \vec{r}_{j}}\Phi_{\alpha\gamma}(\vec{r}_{j},\vec{r}_{l})\,\delta(\vec{p}-\vec{p}_{j})\,\delta(\vec{r}-\vec{r}_{j}),\\
  \vec{F}_{\alpha\xi}(\vec{p},\vec{r})&=\sum_{j,s}^{N_{\alpha}N_{\xi}}\frac{\partial }{\partial \vec{r}_{j}}\Phi_{\alpha\xi}(\vec{r}_{j},\vec{R}_{s})\,\delta(\vec{p}-\vec{p}_{j})\,\delta(\vec{r}-\vec{r}_{j})
 \end{align*}
 are the microscopic densities of interaction forces between ion species $\alpha$ and $\gamma$  and ion species $\alpha$  and particles $\xi$ of the  porous matrix in the space of momentum and coordinates,
 \[
  \vec{E}(\vec{p},\vec{r};t)=\sum_{j}^{N_{\alpha}}\vec{E}(\vec{r}_{j};t)\,\delta(\vec{p}-\vec{p}_{j})\,\delta(\vec{r}-\vec{r}_{j})
 \]
 is the microscopic density of the electric field created by ions of the type $\alpha$ in the space of momentum and coordinates. The electric field $\vec{E}(\vec{r})$ is created by the ions of the solution according to the microscopic Lorentz-Maxwell equation:
 \[
 \frac{\partial }{\partial \vec{r}}\cdot \vec{E}(\vec{r})=\sum_{\alpha}Z_{\alpha}e\hat{n}_{\alpha}(\vec{r}).
 \]
  The action of the Liouville operator on $\hat{\varepsilon}_{\rm int}(\vec{r})$ and on $\hat{n}_{\xi}(\vec{R})$ gives the following result:
 \begin{align*}
  \dot{\hat{\varepsilon}}_{\rm int}(\vec{r})&=\ri L_{N}\hat{\varepsilon}_{\rm int}(\vec{r})=\sum_{\alpha\gamma}\sum_{j=1}^{N_{\alpha}}\frac{\vec{p}_{j}}{m_{\alpha}}\cdot \vec{F}_{\alpha\gamma}(\vec{r}_{j})\,\delta(\vec{r}-\vec{r}_{j})  \\
   &-\frac{\partial }{\partial \vec{r}}\cdot\sum_{\alpha\gamma}\sum_{j=1,l=1}^{N_{\alpha}N_{\gamma}}\frac{\vec{p}_{j}}{m_{\alpha}} \Phi_{\alpha\gamma}(\vec{r}_{j},\vec{r}_{l})\,\delta(\vec{r}-\vec{r}_{j})
   + \sum_{\alpha}\sum_{j=1,l=1}^{N_{\alpha},N_{\xi}}\frac{\vec{P}_{l}}{m_{\xi}}\cdot \vec{F}_{\alpha\xi}(\vec{r}_{j},\vec{R}_{l})\,\delta(\vec{r}-\vec{r}_{j}), \\
  \dot{\hat{n}}_{\xi}(\vec{R})&=\ri L_{N}\hat{n}_{\xi}(\vec{R})=-\frac{1}{m_{\xi}}\frac{\partial }{\partial \vec{R}}\cdot \hat{\vec{P}}_{\xi}(\vec{R}),
 \end{align*}
 where $\hat{\vec{P}}_{\xi}(\vec{R})$ is the microscopic momentum density of the particles of the porous matrix, $\vec{F}_{\alpha\gamma}(\vec{r}_{j})=\sum_{l=1}^{N_{\gamma}}\frac{\partial }{\partial \vec{r}_{l}}\Phi_{\alpha\gamma}(\vec{r}_{j},\vec{r}_{l})$ are the forces acting on the $j$-th ion of the $\alpha$-type from the ions of the $\gamma$-type; $\vec{F}_{\alpha\xi}(\vec{r}_{j},\vec{R}_{l})=\frac{\partial }{\partial \vec{R}_{l}}\Phi_{\alpha\xi}(\vec{r}_{j},\vec{R}_{l})$ are the forces acting on the ion of the type $\alpha$ from the side of the porous matrix particle $\xi$.
 The transport equations (\ref{eq22.9})--(\ref{eq22.9b}) describe non-Markovian processes of ion transport in the system ``ionic solution -- porous medium'', taking into account the dynamics of the porous matrix particles.
 The transfer kernels $\varphi_{nn}^{\alpha\gamma}(x,x';t,t')$, $\varphi_{n\varepsilon}^{\alpha}(x,\vec{r}';t,t')$,
 $\varphi_{nn}^{\alpha\xi}(x,\vec{R}';t,t')$,
 $\varphi_{\varepsilon\varepsilon}(\vec{r},\vec{r}';t,t')$,
 $\varphi_{\varepsilon n}^{\xi}(\vec{r},\vec{R}';t,t')$, $\varphi_{nn}^{\xi\xi}(\vec{R},\vec{R}';t,t')$ have the following structure:
 \begin{equation}\label{e.401}
  \varphi_{AB}(t,t')=\big\langle(1-P(t))\hat{A}\,T_{\rm rel}(t,t')(1-P(t'))\hat{B}\big\rangle_{\rm rel}^{t},
 \end{equation}
 where $\{A,B\}=\{\hat{n}_{\alpha}(x),\, \hat{\varepsilon}_{\rm int}(\vec{r}), \hat{n}_{\xi}(\vec{R})\}$.
 The transfer kernel data are associated with the corresponding generalized transfer coefficients, in particular:
 \begin{equation}\label{e.402}
  \varphi_{nn}^{\xi\xi}(\vec{R},\vec{R}';t,t')
  =
  \frac{\partial }{\partial \vec{R}}\cdot D_{nn}^{\xi\xi}(\vec{R},\vec{R}';t,t')\cdot\frac{\partial }{\partial \vec{R}'},
 \end{equation}
 where
 \begin{equation}\label{e.403}
  D_{nn}^{\xi\xi}(\vec{R},\vec{R}';t,t')=\frac{1}{m_{\xi}^{2}}
  \big\langle (1-P(t))\,\hat{\vec{P}}_{\xi}(\vec{R})\,T_{\rm rel}(t,t')(1-P(t'))\hat{\vec{P}}_{\xi}(\vec{R}')\big\rangle_{\rm rel}^{t}
 \end{equation}
 is the generalized diffusion coefficient of the particles of the porous matrix. For the ionic subsystem:
 \begin{align}\label{e.404}
  \varphi_{nn}^{\alpha\gamma}(x,x';t,t')&=-\frac{1}{m_{\alpha}m_{\gamma}}\frac{\partial }{\partial \vec{r}}\cdot D_{nn}^{\alpha\gamma}(x,x';t,t')\cdot \frac{\partial }{\partial \vec{r}'} \nonumber\\
  &+\frac{1}{m_{\alpha}}\frac{\partial}{\partial\vec{r}}\cdot\bigg\langle (1-P(t))\,\hat{\vec{j}}_{\alpha}(x)\,T_{\rm rel}(t,t')(1-P(t'))
  \bigg(\sum_{\chi}\vec{F}_{\gamma\chi}(x')+Z_{\gamma}e\vec{E}(x';t')\bigg)\bigg\rangle^{t'}_{\rm rel}\cdot \frac{\partial }{\partial \vec{p}'} \nonumber\\
  &+\frac{\partial }{\partial \vec{p}}\cdot\bigg\langle (1-P(t))\bigg(\sum_{\chi}\vec{F}_{\alpha\chi}(x)+Z_{\alpha}e\vec{E}(x;t)\bigg)T_{\rm rel}(t,t')(1-P(t'))
  \hat{\vec{j}}_{\gamma}(x')\bigg\rangle^{t'}_{\rm rel} \frac{1}{m_{\gamma}}\cdot \frac{\partial }{\partial \vec{r}'} \nonumber\\
  &-\frac{\partial }{\partial \vec{p}}\cdot\bigg\langle (1-P(t))\bigg(\sum_{\chi}\vec{F}_{\alpha\chi}(x)+Z_{\alpha}e\vec{E}(x;t)\bigg)T_{\rm rel}(t,t')(1-P(t')) \nonumber\\
  &\times\bigg(\sum_{\chi'}\vec{F}_{\gamma\chi'}(x')+Z_{\gamma}e\vec{E}(x';t')\bigg)\bigg\rangle^{t'}_{\rm rel}\cdot \frac{\partial }{\partial \vec{p}'},
 \end{align}
 where
 \begin{equation}       \label{e.405}
  D_{nn}^{\alpha\gamma}(x,x';t,t')=\big\langle (1-P(t))\hat{\vec{j}}_{\alpha}(x)T_{\rm rel}(t,t')(1-P(t'))\hat{\vec{j}}_{\gamma}(x')\big\rangle^{t'}_{\rm rel}
 \end{equation}
 are the generalized diffusion coefficients of ions in the space of momentum and coordinates, and for $\alpha=\gamma$ $D_{nn}^{++}(x,x';t,t')$, $D_{nn}^{--}(x,x';t,t')$ are the generalized diffusion coefficients of positively and negatively charged ions,
 while $D_{nn}^{+-}(x,x';t,t')$ is their mutual diffusion.
 $\sum_{\chi}\vec{F}_{\alpha\chi}(x)=\sum_{\gamma}\vec{F}_{\alpha\gamma}(x)+\vec{F}_{\alpha\xi}(x)$ is the sum of the microscopic forces in the space of momentum and space acting on the ion of the type $\alpha$ from the ions of the type $\gamma$ and the particles $\xi$ of the porous medium.
 The next two terms in~(\ref{e.404}) describe the momentum-space and time correlations between the forces $\vec{F}_{\alpha\chi}(x)$, $\chi=\{\gamma,\xi\}$, the ion electric field $\vec{E}(x;t)$ and the ion momentum density $\hat{\vec{j}}_{\gamma}(x')$.
 The fourth term in~(\ref{e.404}) describes the momentum-space and time correlations between the forces $\vec{F}_{\alpha\chi}(x)$,
 $\chi=\{\gamma,\xi\}$ and the ion electric field $\vec{E}(x;t)$,
 in particular the term
 \[
  -\frac{\partial }{\partial \vec{p}}\cdot\big\langle (1-P(t))Z_{\alpha}e\vec{E}(x;t)T_{\rm rel}(t,t')(1-P(t'))
 Z_{\gamma}e\vec{E}(x';t')\big\rangle^{t'}_{\rm rel}\cdot \frac{\partial }{\partial \vec{p}'}
 \]
 corresponds to the polarization processes created by the electric field of the ions. At the same time, the term
 \[
  -\frac{\partial }{\partial \vec{p}}\cdot\bigg\langle (1-P(t))\sum_{\chi}\vec{F}_{\alpha\chi}(x)T_{\rm rel}(t,t')(1-P(t'))
 \sum_{\chi'}\vec{F}_{\gamma\chi'}(x')\bigg\rangle^{t'}_{\rm rel}\cdot \frac{\partial }{\partial \vec{p}'},
 \]
 contains
 \begin{equation}\label{e.406}
  \zeta_{FF}^{\alpha\gamma}(x,x';t,t')
  =
  \bigg\langle (1-P(t))\sum_{\chi}\vec{F}_{\alpha\chi}(x)T_{\rm rel}(t,t')(1-P(t'))
  \sum_{\chi'}\vec{F}_{\gamma\chi'}(x')\bigg\rangle^{t'}_{\rm rel}
 \end{equation}
 which is the generalized friction coefficient of ions of the $\alpha$-type and $\gamma$-type taking into account the interaction $\vec{F}_{\alpha\xi}$ between ions and particles of the porous medium.
 Cross-correlation functions
 \[
  -\frac{\partial }{\partial \vec{p}}\cdot\bigg\langle (1-P(t))\sum_{\chi}\vec{F}_{\alpha\chi}(x)T_{\rm rel}(t,t')(1-P(t'))
 Z_{\gamma}e\vec{E}(x';t')\bigg\rangle^{t'}_{\rm rel}\cdot \frac{\partial }{\partial \vec{p}'},
 \]
 describe the interaction between the microscopic forces $\vec{F}_{\alpha\xi}$ and the electric field $\vec{E}(x';t')$ created by the ions.
 Also of interest are the cross-correlation functions:
 \[
  \frac{1}{m_{\alpha}}\frac{\partial }{\partial \vec{r}}\cdot\big\langle (1-P(t))\,
  \hat{\vec{j}}_{\alpha}(x)T_{\rm rel}(t,t')(1-P(t'))
  Z_{\gamma}e\vec{E}(x';t')\big\rangle^{t'}_{\rm rel}\cdot \frac{\partial }{\partial \vec{p}'},
 \]
 which can reflect the influence of the electric field created by ions on their momentum density $\hat{\vec{j}}_{\gamma}(x')$ in the space of momentum and coordinates with evolution in time.
 The transport kernel~(\ref{e.404}) as we can see has a structure of the Fokker--Planck type with the corresponding derivatives $\frac{\partial }{\partial \vec{r}}$ $\frac{\partial }{\partial \vec{p}'}$ in the space of momentum and coordinates.

 In conclusion, equation~(\ref{eq22.9}) describes non-Markovian processes of ion transport by diffusion, friction, electric polarization, and mutual diffusion of ions with particles of a porous medium.
 The transport equation~(\ref{eq22.9a}) for $\langle \hat{\varepsilon}_{\rm int}(\vec{r}) \rangle^{t}$ mainly describes heat-conducting processes for ions in a given system.
 By the method of moments, by projecting the first equation (\ref{eq22.9}) onto the moments $\big(1,\vec{p},\frac{p^{2}}{2m_{\alpha}}\big)$ of the non-equilibrium single-particle distribution function $\langle \hat{n}_{\alpha}(x) \rangle^{t}$, together with equation (\ref{eq22.9a}), we can obtain a system of generalized hydrodynamics equations for the ionic subsystem with the corresponding generalized coefficients of viscosity, thermal conductivity, and thermal diffusion.
 The resulting system of transport equations~(\ref{eq22.9})--(\ref{eq22.9b}) can describe both fast and slow non-Markovian processes of ion transport in a porous medium.
 In this model, we do not specify the Hamiltonian $H_{\xi}$ of the porous medium.
 However, such porous materials, as we have already noted, can be electrode organic and inorganic composite materials, membrane structures for water purification and solution separation, biological membrane structures, and others, in which components (ions, molecules, macromolecules) can have charge, dipole, quadrupole, and magnetic moments, which must be taken into account in $H_{\xi}$.

 In the next section, we consider the subdiffusion model for particles of a porous medium.

 \section{Subdiffusion processes of particles of a porous medium}

 As a result of the interaction of ions with particles of a porous medium over a long time, its structure changes.
 To describe these processes, we assume that the non-equilibrium state of particles of a porous medium is close to the equilibrium state.
 Therefore, we introduce $\delta\mu_{\xi}(\vec{R};t)=\mu_{\xi}(\vec{R};t)-\mu_{\xi}$ as the deviation of the non-equilibrium value of the chemical potential of particles of a porous matrix from its equilibrium value $\mu_{\xi}$.
 We consider such fluctuations to be small.
 Then, in a linear approximation, for $\delta\mu_{\xi}(\vec{R};t)$, the relevant distribution function is given as:
 \begin{equation}\label{e.00211}
  \varrho_{\rm rel}\big(x^{N_{\alpha}},X^{N_{\xi}};t\big)
  =\varrho_{\rm rel}^{(i)}\big(x^{N_{\alpha}},X^{N_{\xi}};t\big)
  \bigg[1-\sum_{\xi}\int \rd\vec{R}\,\beta\delta\mu_{\xi}(\vec{R};t)\,\hat{n}_{\xi}(\vec{R})\bigg],
 \end{equation}
 where
 \begin{align}\label{e.00210}
  \rho_{\rm rel}^{(i)}\big(x^{N_{\alpha}},X^{N_{\xi}};t\big)
  &=\exp\bigg[-\Phi^{(i)}(t)-\int \rd\vec{r}\,\beta(\vec{r},t)\,\hat{\varepsilon}_{\rm int}(\vec{r}) \nonumber\\
  &-\sum_{\alpha}\int \rd x \, a_{\alpha}(x,t)\,\hat{n}_{\alpha}(x)- \beta(H_{\xi}-\mu_{\xi}N_{\xi})\bigg].
 \end{align}
 Using the self-consistency condition $\langle\hat{n}_{\xi}(\vec{R})\rangle^{t}= \langle\hat{n}_{\xi}(\vec{R})\rangle^{t}_{\rm rel}$, we can determine the following fluctuations:
 \begin{equation}\label{e.500}
  \beta\delta\mu_{\xi}(\vec{R};t)=-\int \rd\vec{R}'\,\delta\big\langle\hat{n}_{\xi}(\vec{R}')\big\rangle^{t}
  \big[\tilde{\Phi}_{nn}^{-1}(\vec{R}',\vec{R};t)\big]^{\xi\xi},
 \end{equation}
 where
 \[
  \delta\big\langle\hat{n}_{\xi}(\vec{R}')\big\rangle^{t}
  =
  \big\langle\hat{n}_{\xi}(\vec{R}')\big\rangle^{t}
  -
  \big\langle\hat{n}_{\xi}(\vec{R}')\big\rangle^{(i)}_{\rm rel},
 \]
 $\langle\hat{n}_{\xi}(\vec{R}')\rangle^{(i)}_{\rm rel}=\int \rd\Gamma_{N}\,\hat{n}_{\xi}(\vec{R}')\,\rho_{\rm rel}^{(i)}(x^{N_{\alpha}},X^{N_{\xi}};t)$, $\big[\tilde{\Phi}_{nn}^{-1}(\vec{R}',\vec{R};t)\big]^{\xi\xi}$ are the elements of the inverse matrix to the matrix of density-density correlation functions of the particles of the porous medium:
 \begin{equation}\label{e.501}
  \Phi_{nn}^{\xi\xi}(\vec{R},\vec{R}';t)=\int \rd\Gamma_{N}\,\hat{n}_{\xi}(\vec{R})\,\hat{n}_{\xi}(\vec{R}')\,\rho_{\rm rel}^{(i)}\big(x^{N_{\alpha}},X^{N_{\xi}};t\big).
 \end{equation}
 $\big[\tilde{\Phi}_{nn}^{-1}(\vec{R}',\vec{R};t)\big]^{\xi\xi}$ is determined from the integral equation:
 \[
  \int \rd\vec{R}''\,\big[\tilde{\Phi}_{nn}^{-1}(\vec{R},\vec{R}'';t)\big]^{\xi\xi}
  \Phi_{nn}^{\xi\xi}(\vec{R}'',\vec{R}';t)
  =\delta(\vec{R}-\vec{R}').
 \]
 Next, we substitute (\ref{e.500}) into (\ref{e.00211}), as a result for $\varrho_{\rm rel}\big(x^{N_{\alpha}},X^{N_{\xi}};t\big)$ we obtain:
 \begin{align}\label{e.502}
  \varrho_{\rm rel}\big(x^{N_{\alpha}},X^{N_{\xi}};t\big)
  &=
  \varrho_{\rm rel}^{(i)}\big(x^{N_{\alpha}},X^{N_{\xi}};t\big) \nonumber\\
  &\times
  \bigg[1+\sum_{\xi}\int \rd\vec{R} \int \rd\vec{R}'
  \delta \langle\hat{n}_{\xi}(\vec{R})\rangle^{t}\big[\tilde{\Phi}_{nn}^{-1}(\vec{R},\vec{R}';t)\big]^{\xi\xi}\hat{n}_{\xi}(\vec{R}')\bigg].
 \end{align}
 The approximation for $\varrho_{\rm rel}\big(x^{N_{\alpha}},X^{N_{\xi}};t\big)$ (\ref{e.502}) makes it possible to exclude the parameter $\beta\delta\mu_{\xi}(\vec{R};t)$ from the system of transport equations~(\ref{eq22.9})--(\ref{eq22.9b}).
 In addition, using the self-consistency condition $\langle \hat{n}_{\alpha}(x)\rangle^{t}=\langle \hat{n}_{\alpha}(x)\rangle^{t}_{\rm rel}$,
 we can determine~\cite{Zubarev1991412,Tokarchuk1998687} the parameters $a_{\alpha}(x;t)=-\ln \frac{\langle \hat{n}_{\alpha}(x)\rangle^{t}}{u_{\alpha}(\vec{r},t)}$,
 where $u_{\alpha}(\vec{r},t)$ satisfies the corresponding integral equation:
 \begin{equation}\label{e.502aa}         
  u_{\alpha}(\vec{r},t)=\int \frac{\rd\vec{r}'^{N_{\alpha}}}{(N_{\alpha}-1)!} \frac{\rd\vec{R}^{N_{\xi}}}{(N_{\xi})!}
         \exp \left[U\big(\vec{r},\vec{r}'^{N_{\alpha}-1},\vec{R}^{N_{\xi}};t\big)\right]
         \prod_{\gamma,j=1}^{N_{\gamma}}\frac{n_{\gamma}(\vec{r}'_{j};t)}{u_{\gamma}(\vec{r}'_{j};t)},
 \end{equation}
 here, the integration over $\rd\vec{r}'$ coordinates of ions is marked by a dash to highlight $u_{\alpha}(\vec{r},t)$,
 \begin{equation}\label{e.502a1}
  n_{\gamma}(\vec{r};t)=\int \rd\vec{p}\langle \hat{n}_{\gamma}(x) \rangle^{t}
 \end{equation}
 is the non-equilibrium average value of the ion density of the species $\gamma$,
 \begin{align}\label{e.502a2}
  U\big(\vec{r}, \vec{r}'^{N_{\alpha}-1},\vec{R}^{N_{\xi}};t\big)
    &=
   \Phi(t)+\sum_{\alpha}\left[\sum_{\gamma}\sum_{j,l}^{N_{\alpha},N_{\gamma}}\frac{1}{2} \Phi_{\alpha\gamma}(\vec{r}'_{j},\vec{r}'_{l})
   +\sum_{j,l}^{N_{\alpha},N_{\xi}}\Phi_{\alpha \xi}(\vec{r}'_{j},\vec{R}_{l})\right]\beta(\vec{r}'_{j};t) \nonumber\\
   &+\beta\left[H_{\xi}-\sum_{\xi}\int \rd\vec{R}\,\mu_{\xi}(\vec{R};t)\,\hat{n}_{\xi}(\vec{R})\right].
  \end{align}
 Then, the system of transport equations~(\ref{eq22.9})--(\ref{eq22.9b}) has the form:
 \begin{align}\label{eq22.99}
  \frac{\partial }{\partial t}\langle \hat{n}_{\alpha}(x) \rangle^{t}&=\langle \dot{\hat{n}}_{\alpha}(x) \rangle^{t}_{\rm rel}
  -\sum_{\gamma}\int \rd x'\int_{-\infty}^{t}\re^{\varepsilon (t'-t)}\varphi_{nn}^{\alpha\gamma}(x,x';t,t')\ln \frac{\langle \hat{n}_{\gamma}(x') \rangle^{t'}}{u_{\gamma}(\vec{r}',t')}\,\rd t' \nonumber\\
  &+\int \rd\vec{r}'\int_{-\infty}^{t}\re^{\varepsilon (t'-t)}\varphi_{n\varepsilon}^{\alpha}(x,\vec{r}';t,t')\,\beta (\vec{r}',t')\,\rd t' \nonumber\\
  &+\int \rd\vec{R}'\int_{-\infty}^{t}\re^{\varepsilon (t'-t)}\bar{\varphi}_{nn}^{\alpha\xi}(x,\vec{R}';t,t')\,\delta\langle\hat{n}_{\xi}(\vec{R}')\rangle^{t'}\rd t',
 \end{align}
 \begin{align}\label{eq22.99a}
  \frac{\partial }{\partial t}\langle \hat{\varepsilon}_{\rm int}(\vec{r}) \rangle^{t}&=\langle \dot{\hat{\varepsilon}}_{\rm int}(\vec{r}) \rangle^{t}_{\rm rel}
  -\sum_{\gamma}\int \rd x'\int_{-\infty}^{t}\re^{\varepsilon (t'-t)}\varphi_{\varepsilon n}^{\gamma}(\vec{r},x';t,t')\ln\frac{\langle \hat{n}_{\gamma}(x') \rangle^{t'}}{u_{\gamma}(\vec{r}',t')}\,\rd t'\nonumber\\
  &+ \int \rd\vec{r}'\int_{-\infty}^{t}\re^{\varepsilon (t'-t)}\varphi_{\varepsilon\varepsilon}(\vec{r},\vec{r}';t,t')\,\beta(\vec{r}',t')\,\rd t'\nonumber\\
  &+ \int \rd\vec{R}'\int_{-\infty}^{t}\re^{\varepsilon (t'-t)}\bar{\varphi}_{\varepsilon n}^{\xi}(\vec{r},\vec{R}';t,t')\delta \langle\hat{n}_{\xi}(\vec{R}')\rangle^{t'}\rd t',
 \end{align}
 \begin{align}    \label{eq22.99b}
  \frac{\partial }{\partial t}\delta\langle \hat{n}_{\xi}(\vec{R}) \rangle^{t}
  &=-\sum_{\gamma}\int \rd x'\int_{-\infty}^{t}\re^{\varepsilon (t'-t)}\varphi_{nn}^{\xi\gamma}(\vec{R},x';t,t')\ln\frac{\langle \hat{n}_{\gamma}(x') \rangle^{t'}}{u_{\gamma}(\vec{r}',t')}\,\rd t' \nonumber\\
  &+ \int \rd\vec{r}'\int_{-\infty}^{t}\re^{\varepsilon (t'-t)}\varphi_{n\varepsilon}^{\xi}(\vec{R},\vec{r}';t,t')\,\beta (\vec{r}',t')\,\rd t' \nonumber\\
  &+ \int \rd\vec{R}'\int_{-\infty}^{t}\re^{\varepsilon (t'-t)}\bar{\varphi}_{nn}^{\xi\xi}(\vec{R},\vec{R}';t,t')\delta \langle\hat{n}_{\xi}(\vec{R}')\rangle^{t'}\rd t',
 \end{align}
 which contains renormalized generalized transport kernels (diffusion coefficient of porous matrix particles) through the structure functions $\big[\tilde{\Phi}_{nn}^{-1}(\vec{R}'',\vec{R}';t')\big]^{\xi\xi}$:
 \begin{equation}    \label{e.504}
  \bar{\varphi}_{nn}^{\xi\xi}(\vec{R},\vec{R}';t,t')=\int \rd\vec{R}'' \frac{\partial }{\partial \vec{R}}\cdot D_{nn}^{\xi\xi}(\vec{R},\vec{R}'';t,t')\cdot\frac{\partial }{\partial \vec{R}''}\big[\tilde{\Phi}_{nn}^{-1}(\vec{R}'',\vec{R}';t')\big]^{\xi\xi},
 \end{equation}
 \begin{equation}\label{e.505}
  \bar{\varphi}_{nn}^{\alpha\xi}(x,\vec{R}';t,t')=\int \rd\vec{R}''\big\langle (1-P(t))\hat{n}_{\alpha}(x)T_{\rm rel}(t,t')(1-P(t'))\hat{n}_{\xi}(\vec{R}'')\big\rangle_{\rm rel}^{t'}
  \big[\tilde{\Phi}_{nn}^{-1}(\vec{R}'',\vec{R}';t')\big]^{\xi\xi},
 \end{equation}
 \begin{equation}    \label{e.506}
  \bar{\varphi}_{\varepsilon n}^{\xi}(\vec{r},\vec{R}';t,t')
  =
  \int \rd\vec{R}'' \big\langle (1-P(t))\hat{\varepsilon}_{\rm int}(\vec{r})T_{\rm rel}(t,t')(1-P(t'))\hat{n}_{\xi}(\vec{R}'')\big\rangle_{\rm rel}^{t'}
 [\tilde{\Phi}_{nn}^{-1}(\vec{R}'',\vec{R}';t')]^{\xi\xi}.
 \end{equation}
 In this case, the averaging operation in the transport kernels of the system of transport equations~(\ref{eq22.99})--(\ref{eq22.99b}) is performed with the relevant distribution function (\ref{e.502}).

 To describe the subdiffusion processes of particles of a porous medium when interacting with an ionic solution, we use the fractional calculus based on the works~\cite{Kostrobij20184099,Kostrobij2021103304}.
 To reveal the time multifractality in the generalized diffusion equation of particles of a porous medium~(\ref{eq22.99b}),
 we use the approximation for the generalized diffusion coefficient
 \begin{equation}\label{e.507}
  D_{nn}^{\xi\xi}(\vec{R},\vec{R}'';t,t')=W_{\xi\xi}(t,t')\,\bar{D}_{nn}^{\xi\xi}(\vec{R},\vec{R}''),
 \end{equation}
 where the function $W_{\xi\xi}(t,t')$ can be defined as a memory function with time retardation.
 Further, as in the works~\cite{Kostrobij20184099,Kostrobij2021103304},
 we use the Fourier transform to equation~(\ref{eq22.99b}),
 choosing the frequency dependence of the memory function in the form
 \begin{equation}\label{e.508}
  W_{\xi\xi}(\omega)=\frac{(\ri\omega)^{1-\nu}}{1+\ri\omega \tau_{\xi}},\quad  0<\nu \leqslant 1,
 \end{equation}
 with the introduction of the relaxation time $\tau_{\xi}$, which characterizes the subdiffusion processes of particles in a porous medium.
 The next step is to use the Fourier transform to fractional derivatives of the functions:
 \begin{equation}\label{eq:2.1096}
  L\big( _{0}D_{t}^{1-\nu}f(t);\ri\omega\big)=(\ri\omega)^{1-\nu} L(f(t);\ri\omega),
 \end{equation}
 where
 \[
  _{0}D_{t}^{1-\nu}f(t)=\frac{1}{\Gamma (\nu)}\frac{\rd}{\rd t}\int_{0}^{t}\frac{f(\tau)}{(t-\tau)^{1-\nu}}\rd\tau,\quad 
  0<\nu\leqslant 1
 \]
 is a fractional Rieman--Liouville derivative and $\Gamma (\nu)$ is the gamma-function.
 Using it, the inverse Fourier transforms in equation~(\ref{eq22.99b}) to the time dependence, taking into account (\ref{e.507}) and (\ref{e.508}),
 gives the generalized diffusion equation for particles of a porous medium of the Cattaneo-type in fractional derivatives:
 \begin{align}\label{e.509}
  \tau_{\xi}\frac{\partial^{2} }{\partial t^{2}}\delta\langle \hat{n}_{\xi}(\vec{R})\rangle^{t}+\frac{\partial }{\partial t}\delta\langle \hat{n}_{\xi}(\vec{R}) \rangle^{t}
  &=
  -\left(1+\tau_{\xi}\frac{\partial }{\partial t}\right)  \nonumber\\
  &\times\sum_{\gamma}\int \rd x'\int_{-\infty}^{t}\re^{\varepsilon (t'-t)}\varphi_{nn}^{\xi\gamma}(\vec{R},x';t,t')\ln \frac{\langle \hat{n}_{\gamma}(x') \rangle^{t'}}{u_{\gamma}(\vec{r}',t')}\,\rd t'\nonumber\\
  &+\left(1+\tau_{\xi}\frac{\partial }{\partial t}\right)
  \int \rd\vec{r}'\int_{-\infty}^{t}\re^{\varepsilon (t'-t)}\varphi_{n\varepsilon}^{\xi}(\vec{R},\vec{r}';t,t')\,\beta(\vec{r}',t')\,\rd t' \nonumber\\
  &+{}_{0}D_{t}^{1-\nu}\int \rd\vec{R}'\int \rd\vec{R}''\frac{\partial }{\partial \vec{R}}\cdot \bar{D}_{nn}^{\xi\xi}(\vec{R},\vec{R}'') \nonumber\\
  &\times
  \frac{\partial }{\partial \vec{R}''}\big[\tilde{\Phi}_{nn}^{-1}(\vec{R}'',\vec{R}';t)\big]^{\xi\xi}\delta \langle\hat{n}_{\xi}(\vec{R}')\rangle^{t}.
  \end{align}
 Taking into account equations (\ref{eq22.99}), (\ref{eq22.99a}) and (\ref{e.509}),
 we obtain a system of equations for the transport of solution ions in a porous medium with a consistent description of the kinetics and hydrodynamics of the liquid subsystem and subdiffusion processes of particles of the porous subsystem.
 The cross-transport kernels $\bar{\varphi}_{nn}^{\alpha\xi}(x,\vec{R}';t,t')$, $\bar{\varphi}_{\varepsilon n}^{\xi}(\vec{r},\vec{R}';t,t')$ describe the dynamic correlations between the particles of the subsystems,
 in particular, $\bar{\varphi}_{nn}^{\alpha\xi}(x,\vec{R}';t,t')$ includes the generalized coefficient of mutual diffusion of solution ions and particles of the porous medium.
 Without taking into account the energy transfer of the interaction of solution ions, this system is significantly simplified:
 \begin{align}\label{eq22.999}
  \frac{\partial }{\partial t}\langle \hat{n}_{\alpha}(x) \rangle^{t}&=\langle \dot{\hat{n}}_{\alpha}(x) \rangle^{t}_{\rm rel}
  -s_{l}\phi\sum_{\gamma}\int \rd x'\int_{-\infty}^{t}\re^{\varepsilon (t'-t)}\bar{\varphi}_{nn}^{\alpha\gamma}(x,x';t,t')\ln\frac{\langle \hat{n}_{\gamma}(x') \rangle^{t'}}{u_{\gamma}(\vec{r}',t')}\rd t' \nonumber\\
  &+ (1-\phi) \int \rd\vec{R}'\int_{-\infty}^{t}\re^{\varepsilon (t'-t)}\bar{\varphi}_{nn}^{\alpha\xi}(x,\vec{R}';t,t')\, \delta \langle\hat{n}_{\xi}(\vec{R}')\rangle^{t'}\rd t',
 \end{align}
 \begin{align}\label{e.5099}
  \tau_{\xi}\frac{\partial^{2} }{\partial t^{2}}\delta\langle \hat{n}_{\xi}(\vec{R}) \rangle^{t}+\frac{\partial }{\partial t}\delta\langle \hat{n}_{\xi}(\vec{R}) \rangle^{t}   &= 
  -\left(1+\tau_{\xi}\frac{\partial }{\partial t}\right) \nonumber\\
  &\times s_{l}\phi\sum_{\gamma}\int \rd x'\int_{-\infty}^{t}\re^{\varepsilon (t'-t)}\varphi_{nn}^{\xi\gamma}(\vec{R},x';t,t')
  \ln \frac{\langle \hat{n}_{\gamma}(x') \rangle^{t'}}{u_{\gamma}(\vec{r}',t')}\,\rd t' \nonumber\\
  &+_{0}D_{t}^{1-\nu}(1-\phi)\int \rd\vec{R}'\int d\vec{R}''\frac{\partial }{\partial \vec{R}}\cdot \bar{D}_{nn}^{\xi\xi}(\vec{R},\vec{R}'')\nonumber\\
  &\times\frac{\partial }{\partial \vec{R}''}\big[\tilde{\Phi}_{nn}^{-1}(\vec{R}'',\vec{R}';t)\big]^{\xi\xi}\delta \langle\hat{n}_{\xi}(\vec{R}')\rangle^{t},
 \end{align}
 where the porosity $\phi$ of the medium and the fraction of filling $s_{l}$ with an ionic solution of the porous medium are taken into account.
 We actually obtain a kinetic equation for the non-equilibrium single-ion distribution function $\langle \hat{n}_{\alpha}(x) \rangle^{t}=f_{\alpha}(x;t)$, which is related to the generalized diffusion equation of the Cattaneo-type in fractional derivatives for particles of the porous medium.
 It is also important to note that in the kinetic equation in $\langle\dot{\hat{n}}_{\alpha}(x)\rangle^{t}_{\rm rel}$ and the transport kernels $\bar{\varphi}_{nn}^{\alpha\gamma}(x,x';t,t')$,
 $\bar{\varphi}_{nn}^{\alpha\xi}(x,\vec{R}';t,t')$ there are contributions from the electric field created by the ions and the gradient of the external electric potential, if present.
 And obviously, it is necessary to add the corresponding Maxwell equations for the averaged electric and magnetic fields to the system of transport equations. These issues require an additional research, which will be carried out in future works.
 We can introduce a generalization to spatial fractality in the Liouville operator $\ri L_{N}(t)$  and in the transport equations by formal substitution according to the works~\cite{Tarasov2004123,Tarasov200517,Tarasov2005011102,Tarasov2010,Kostrobij201963}:
 $\int \rd x\rightarrow \int \rd\mu_{\varsigma}(x)$, $0<\varsigma <1$, $\frac{\partial }{\partial \vec{r}}\rightarrow D_{\vec{r}}^{\varsigma}$, $\frac{\partial }{\partial \vec{p}}\rightarrow D_{\vec{p}}^{\varsigma}$, where $D_{\vec{r}}^{\varsigma}$, $D_{\vec{p}}^{\varsigma}$ are the derivative operators of the Riemann--Liouville type.

 \section{Conclusions}

 A consistent description of kinetic and hydrodynamic processes is applied to the study of ion transport processes in the system ``ionic solution -- porous medium''. 
 A non-equilibrium particle distribution function for the system ``ionic solution -- porous medium'' is obtained using the non-equilibrium statistical operator method. 
 When choosing a distribution function for solving the Liouville equation,
 we assumed that at the initial moment of time, the temperature of the porous subsystem is constant, 
 and the chemical potential of its particles changes due to the interaction between them and the solution ions. 
 The motion of ions was described at the kinetic level in the coordinate and momentum space using $\langle\hat{n}_{\alpha}(x)\rangle^{t}=f_{\alpha}(x;t)$. 
 In this case, 
 the non-equilibrium average value of the interaction energy $\langle \hat{\varepsilon}_{\rm int}(\vec{r})\rangle^{t}$ contains, 
 in addition to the ion-ion interaction, 
 the ion-particle interaction of the matrix, 
 so the non-equilibrium temperature of the ionic solution is a certain effective temperature, 
 which already has a certain contribution from the porous subsystem through diffusion processes.
 In this case, 
 the non-equilibrium properties of the ionic solution are described within the framework of the consistency of kinetics and hydrodynamics, 
 and the particles of the porous medium -- within the framework of the diffusion (subdiffusion) description. 
 That is, 
 the ionic transport of ions in a porous medium is described using the kinetic equation~(\ref{eq22.9}) (taking into account (\ref{e.404}) with the Fokker--Planck type interaction integral in the coordinate and momentum space), 
 and therefore it can be both fast and slow and it also depends on the equation~(\ref{eq22.9a}) of the change in time of the average interaction energy. 
 The resulting system of transport equations~(\ref{eq22.9})--(\ref{eq22.9b}) can describe both fast and slow non-Markovian processes of the ion transport in a porous medium. 
 We analyzed the structure of generalized transport kernels in the obtained system of equations with the separation of the interaction forces between ions and particles of the porous medium and the contributions of the electric field created by solution ions. 
 As a result, 
 in particular, 
 the transport kernels were expressed in terms of generalized diffusion coefficients, 
 friction in coordinate space, 
 and momenta for the ionic subsystem in the Fokker--Planck type collision integral.

 As experimental studies show, 
 in particular in the processes of operation of porous electrodes in batteries, 
 in membrane technologies, 
 as a result of the interaction of ions with particles of the porous medium over a long period of time, 
 its structure changes significantly. 
 As a rule, 
 from the point of view of theoretical description, 
 these changes are recorded at the level of modelling the porosity of the medium $\phi$. 
 It is obvious that the change in the structure of the porous matrix is related to the movement of its particles, 
 which is very slow during the operation time. 
 Therefore, 
 to describe such processes, 
 we considered that the non-equilibrium state of the particles of the porous medium is close to the equilibrium state and its state can be described in a linear approximation by $\delta\mu_{\xi}(\vec{R};t)=\mu_{\xi}(\vec{R};t)-\mu_{\xi}$ --- the deviation of the non-equilibrium value of the chemical potential of the particles of the porous matrix from its equilibrium value $\mu_{\xi}$. 
 In this approximation, 
 with the appropriate transformations, 
 the system of equations~(\ref{eq22.9})--(\ref{eq22.9b}) was written in the form (\ref{eq22.99})--(\ref{eq22.99b}) with the fact that the equation of generalized diffusion of particles of a porous matrix is already closed with respect to $\delta\langle \hat{n}_{\xi}(\vec{R}) \rangle^{t}$.
 Further, 
 having some personal experience in describing subdiffusion processes~\cite{Kostrobij2015154,Grygorchak2017185501,Kostrobij20184099,Kostrobij2021103304,Kostrobij201963} using the fractional calculus technique~\cite{Samko1993,Oldham2006}, 
 the equation~(\ref{eq22.99b}) is written in the Cattaneo form~(\ref{e.509}) in fractional derivatives to describe the processes of subdiffusion of particles in a porous medium. 
 In order to perform numerical calculations in the following works, 
 we presented the system of equations~(\ref{eq22.99}), (\ref{e.509}) in a closed form with an explicit inclusion of the porosity $\phi$ of the medium, 
 but without taking into account the contributions $\langle \hat{\varepsilon}_{\rm int}(\vec{r})\rangle^{t}$.
 We hope that by using the method of moments for correction functions and isolating the Gaussian or Lorentzian form of the time dependence, 
 we will obtain expressions for the transport kernels in terms of the corresponding structural distribution functions and interaction potentials between ions and particles of the porous matrix of a specific model. 
 On their basis, 
 numerical calculations of the corresponding transport equations~(\ref{eq22.99}), (\ref{e.509} will be performed.

\ukrainianpart

\title{Узгоджений опис кінетичних процесів транспорту іонів електроліту в динамічному пористому середовищі}

\author{П. П. Костробій\refaddr{label1}, Б. М. Маркович\refaddr{label1}, О. В. Візнович\refaddr{label1}, М. В. Токарчук\refaddr{label2,label1}}

\addresses{
\addr{label1}Національний університет ``Львівська політехніка'', 79013, м. Львів, вул. С. Бандери, 12
\addr{label2}Інститут фізики конденсованих систем НАН України, 79011 Львів, вул. Свєнціцького, 1
}

\makeukrtitle

\begin{abstract}
\tolerance=3000%
 Узгоджений опис кiнетичних та гiдродинамiчних процесiв застосовано до дослiджень процесiв переносу iонiв у системi ``iонний розчин -- пористе середовище''. Отримано систему рiвнянь для нерiвноважної одноiонної функцiї розподiлу, нерiвноважного середнього значення густини енергiї взаємодiї iонiв розчину  та для нерiвноважного середнього значення густини числа частинок пористого середовища. З використан\-ням методики дробового числення
 отримано узагальнене рiвняння дифузiї типу Кеттано у дробових
  похiдних для опису процесiв субдифузiї частинок пористого середовища.

\keywords кінетичні та гідродинамічні процеси,  нерівноважний статистичний оператор, іонні розчини, пористе середовище, дробове числення, дифузія.

\end{abstract}

\end{document}